\documentclass[dvips]{article}
\usepackage{icrctc07}

\title{Energy and composition sensitivity of geosynchrotron radio emission from EAS}
\shorttitle{Energy and composition with radio}
\authors{T. Huege$^{1}$, R. Ulrich$^{1}$, R. Engel$^{1}$.}
\shortauthors{T. Huege and et al}
\afiliations{$^1$Institut f\"ur Kernphysik, Forschungszentrum Karlsruhe, Postfach 3640, 76021 Karlsruhe, Germany}
\email{tim.huege@ik.fzk.de}

\abstract{We analyse the sensitivity of geosynchrotron radio emission from inclined extensive air showers to the energy and mass of primary cosmic rays. We demonstrate that radio emission measurements at suitable lateral distances can infer both the number of electrons and positrons in the shower maximum and the atmospheric depth of the maximum on a shower-to-shower basis. Alternatively, measurements at a fixed lateral distance but in two different observing frequency bands yield comparable information. An RMS error of 5\% in the determination of the number of electrons and positrons at shower maximum can be achieved. Through the determination of these quantities, geosynchrotron radiation provides access to the energy and mass of primary cosmic rays on a shower-to-shower basis.}

\begin{document}
\maketitle
%Begin the section.

\section{Introduction}

It is our current understanding that radio emission from extensive air showers is dominated by ``geosynchrotron radiation'' \cite{HuegeFalcke2003a} emitted by secondary shower electrons and positrons being deflected in the earth's magnetic field. The geosynchrotron model has by now been implemented in a sophisticated Monte Carlo code called REAS2 \cite{HuegeUlrichEngel2007a,HuegeIcrc2007b}, which itself relies on CORSIKA \cite{HeckKnappCapdevielle1998} for the simulation of the relevant extensive air shower (EAS) properties. In this article, we analyse the sensitivity of REAS2-simulated geosynchrotron radio emission on the primary cosmic ray energy and mass. The determination of these parameters on a shower-to-shower basis is one of the main goals of measuring radio emission from EAS, and we demonstrate that radio measurements of inclined showers indeed provide relatively direct access to these parameters.

\section{Methodology}

Results gathered so far point to inclined air showers with zenith angles above $45^{\circ}$ as particularly promising targets for radio measurements of EAS \cite{PetrovicApelAsch2006}. Inclined air showers have a large radio ``footprint'' \cite{HuegeFalcke2005b} and thus allow radio antennas to be spaced relatively far apart, an important prerequisite for the instrumentation of large effective areas at moderate cost. Concentrating on the energy range relevant to the Pierre Auger Observatory, we have thus performed an analysis of air showers with $60^{\circ}$ and $45^{\circ}$ zenith angles at primary particle energies of $10^{18}$~eV, $10^{19}$~eV and $10^{20}$~eV. For each of these zenith angles and energies, we have simulated 25 iron-induced and 25 proton-induced air showers. 25 gamma-induced air showers per zenith angle were simulated for $10^{18}$~eV and $10^{19}$~eV, but not for $10^{20}$~eV, where pre-showering in the geomagnetic field would have to be taken into account. The simulation chain consisted of a CORSIKA 6.502 run using the QGSJETII-03 and UrQMD1.3.1 interaction models, Argentinian magnetic field, fixed azimuth angle (showers coming from south), $10^{-6}$ optimised thinning, and an observer height 1400~m a.s.l. followed by a REAS2 simulation with antenna locations between 25~m and 925~m from the shower core (in ground-based coordinates); cf.\ also \cite{HuegeUlrichEngel2007a}. For each of the simulated radio events we then determined the peak amplitude of the electric field pulses filtered using idealised rectangle filters from 16 to 32~MHz, 32 to 64~MHz and 64 to 128~MHz, respectively. In the following, we discuss the $60^{\circ}$ zenith angle case. The qualitative behaviour at $45^{\circ}$ is completely analogue. Other shower azimuth angles do not change the qualitative behaviour either.

\section{Signal information content}

\begin{figure*}[!htb]
\begin{minipage}{0.47\textwidth}
\centering
\includegraphics [width=4.80cm,angle=270]{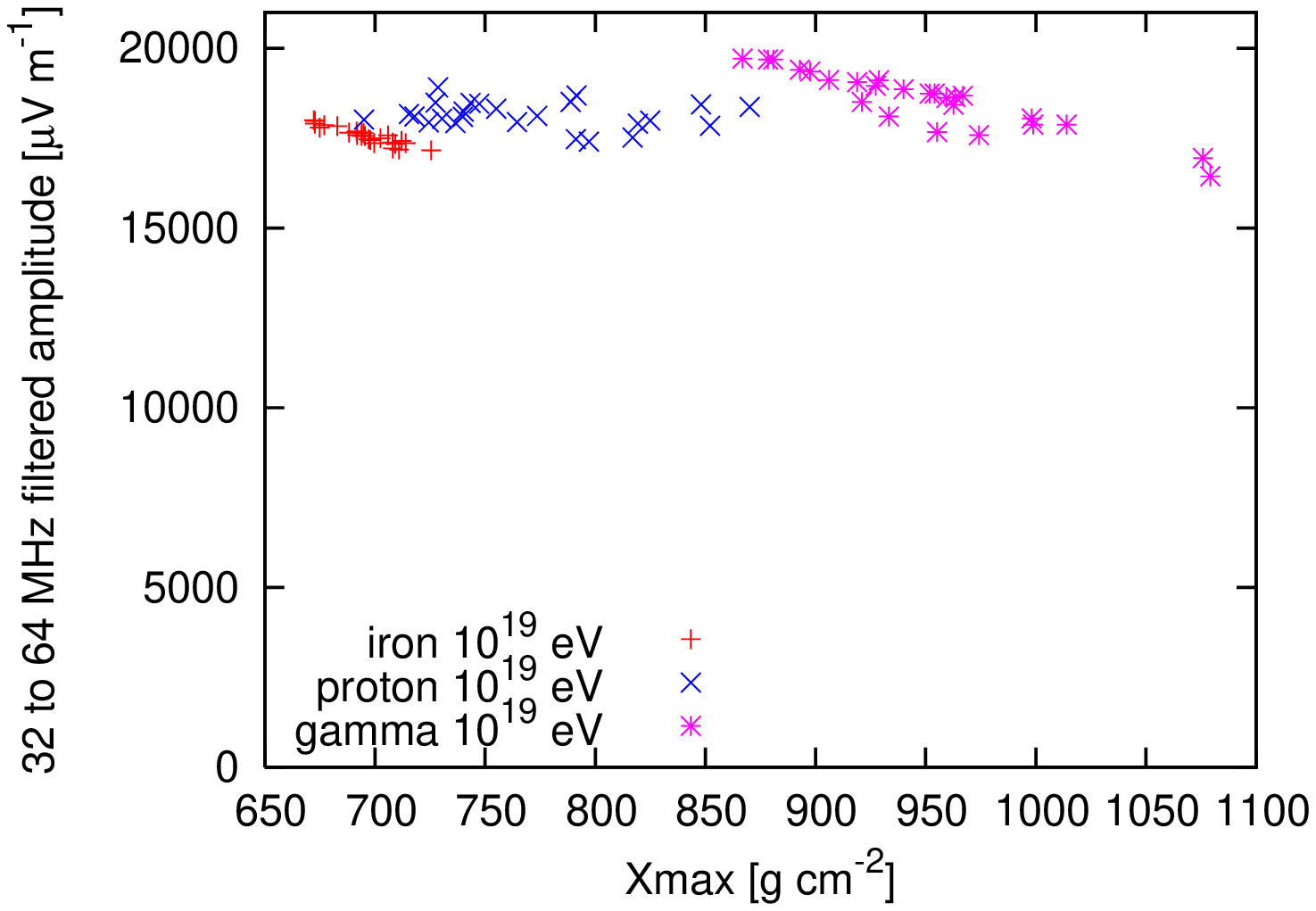}
\caption{Peak radio field strengths of $10^{19}$~eV showers with $60^{\circ}$ zenith angle as measured in the {\em flat region} 275~m north of the shower core.}\label{fig:flat1e19}
\end{minipage}
\hspace{0.05\textwidth}
\begin{minipage}{0.47\textwidth}
\centering
\includegraphics [width=4.80cm,angle=270]{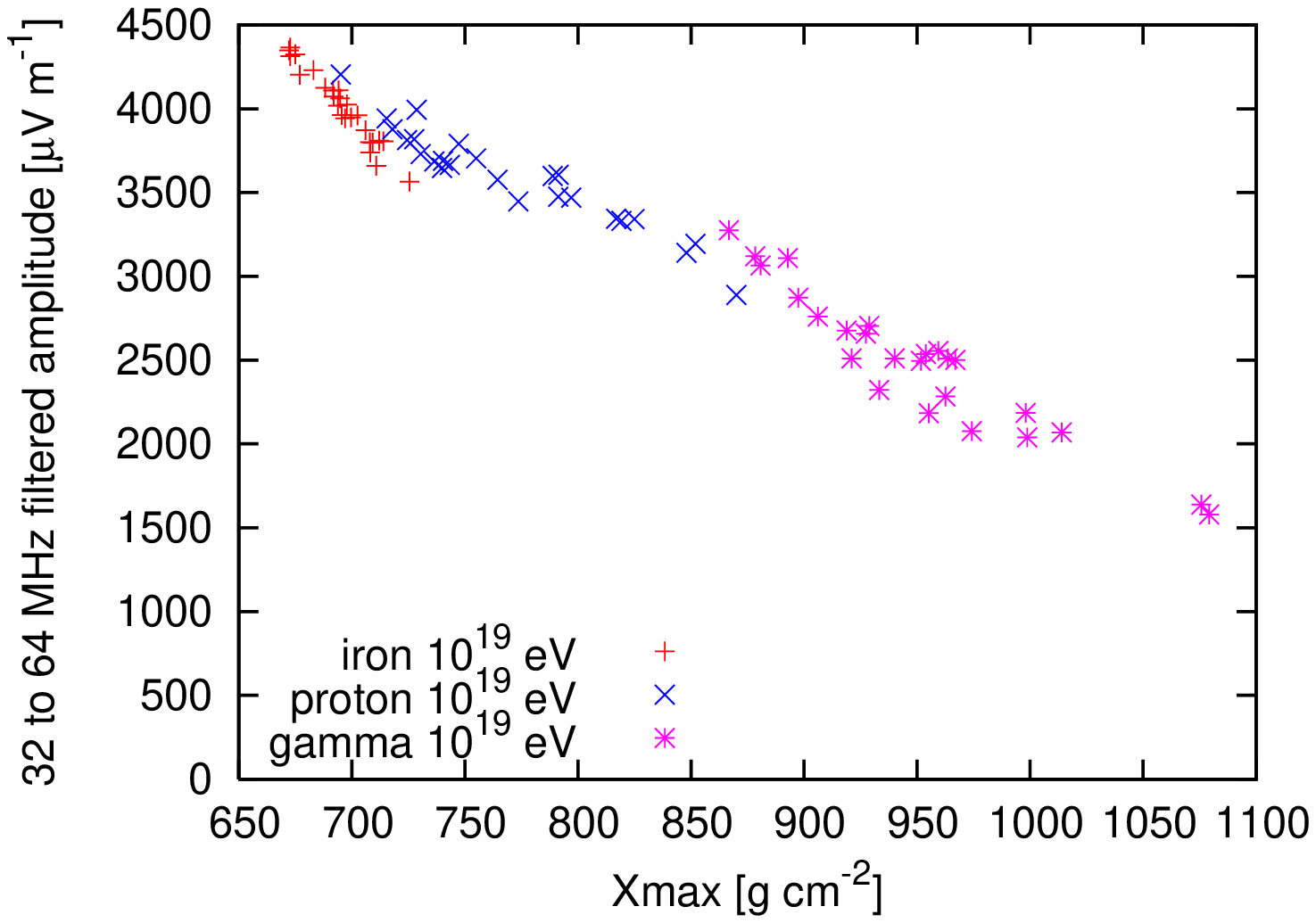}
\caption{Peak radio field strengths of $10^{19}$~eV showers with $60^{\circ}$ zenith angle as measured in the {\em steep region} 725~m north of the shower core.}\label{fig:steep1e19}
\end{minipage}
\end{figure*}

The lateral slope of the radio signal is known to exhibit a dependence on the depth of the shower maximum ($X_{\mathrm{max}}$) and consequently, contains information on the primary particle energy and mass \cite{HuegeFalcke2005b,HuegeIcrc2005a}. To date, however, there had been no detailed investigation how this information content could be exploited in practice.

\begin{figure}[hb]
\centering
\includegraphics [width=6.0cm,angle=270]{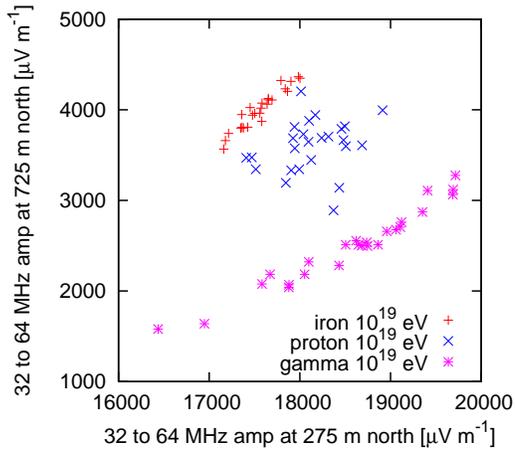}
\caption{The ratio of field strengths in the {\em steep region} vs.\ the {\em flat region} of $10^{19}$~eV showers with $60^{\circ}$ distinguishes between primary particle types.}\label{fig:ratio}
\end{figure}
 
The detailed analysis presented here reveals that for a given shower zenith angle, a suitable lateral distance exists where the radio signal is relatively independent of $X_{\mathrm{max}}$ (hereafter called {\em flat region}), in analogy to the surface detector quantity $S(1000)$ of the Pierre Auger Observatory. This behaviour is illustrated for the $60^{\circ}$ zenith angle case in Fig.\ \ref{fig:flat1e19}: the electric field strength at 275~m north from the shower core is relatively constant regardless of primary particle type and shower $X_{\mathrm{max}}$. In contrast, at a distance of 725~m north as shown in Fig.\ \ref{fig:steep1e19}, there is a clear dependence of signal strength on $X_{\mathrm{max}}$ (hereafter called {\em steep region}).

A combination of measurements in these two regions can therefore differentiate between different types of primaries (Fig.\ \ref{fig:ratio}).

\section{Signal scaling with $N_{\mathrm{max}}$}

\begin{figure*}[!htb]
\begin{minipage}{0.47\textwidth}
\centering
\includegraphics[width=4.80cm,angle=270]{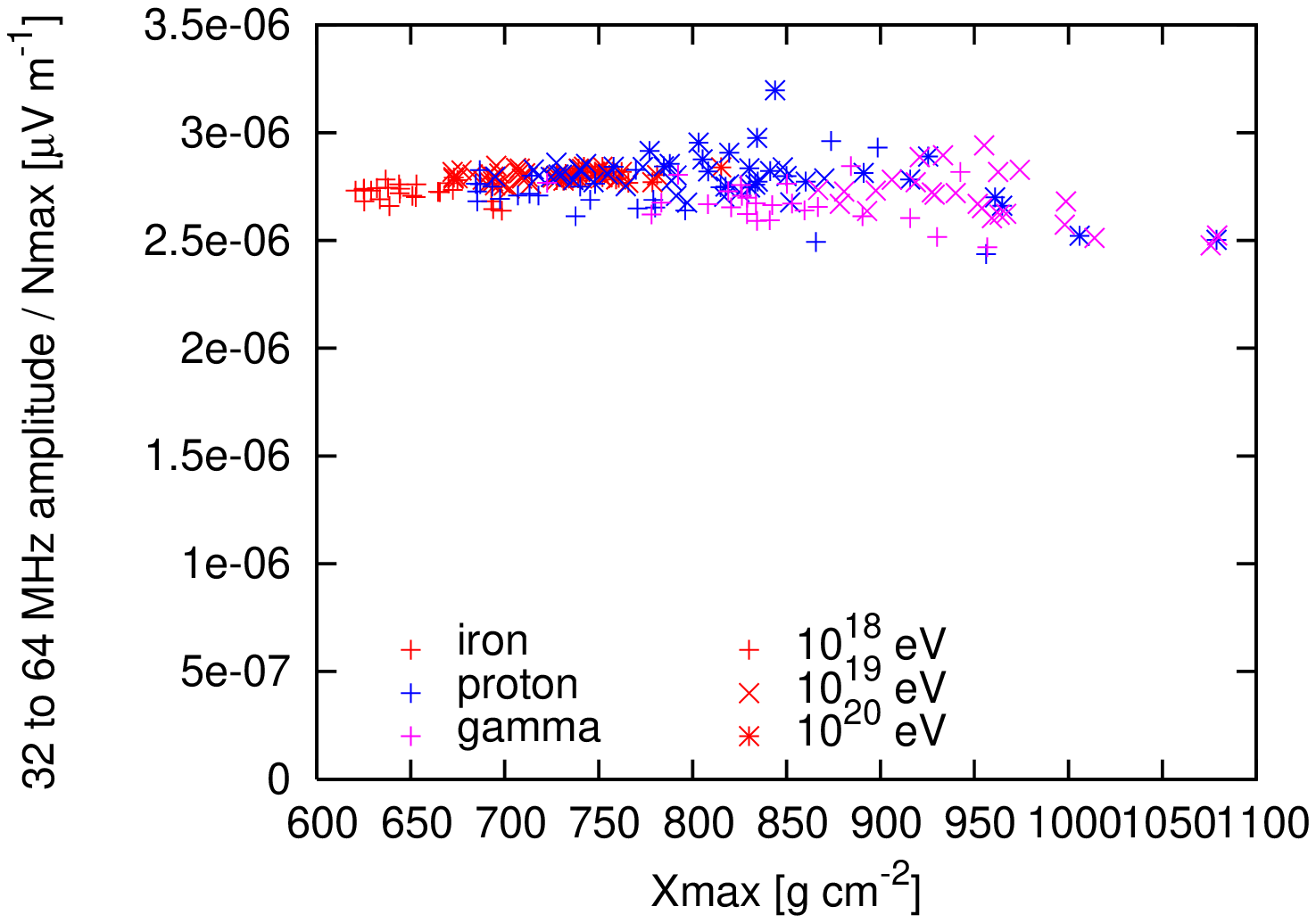}
\caption{In the {\em flat region} at 275~m north, the radio pulse height per $N_{\mathrm{max}}$ over all energies and particle types is approximately constant.}\label{fig:flatnorm}
\end{minipage}
\hspace{0.05\textwidth}
\begin{minipage}{0.47\textwidth}
\centering
\includegraphics[width=4.80cm,angle=270]{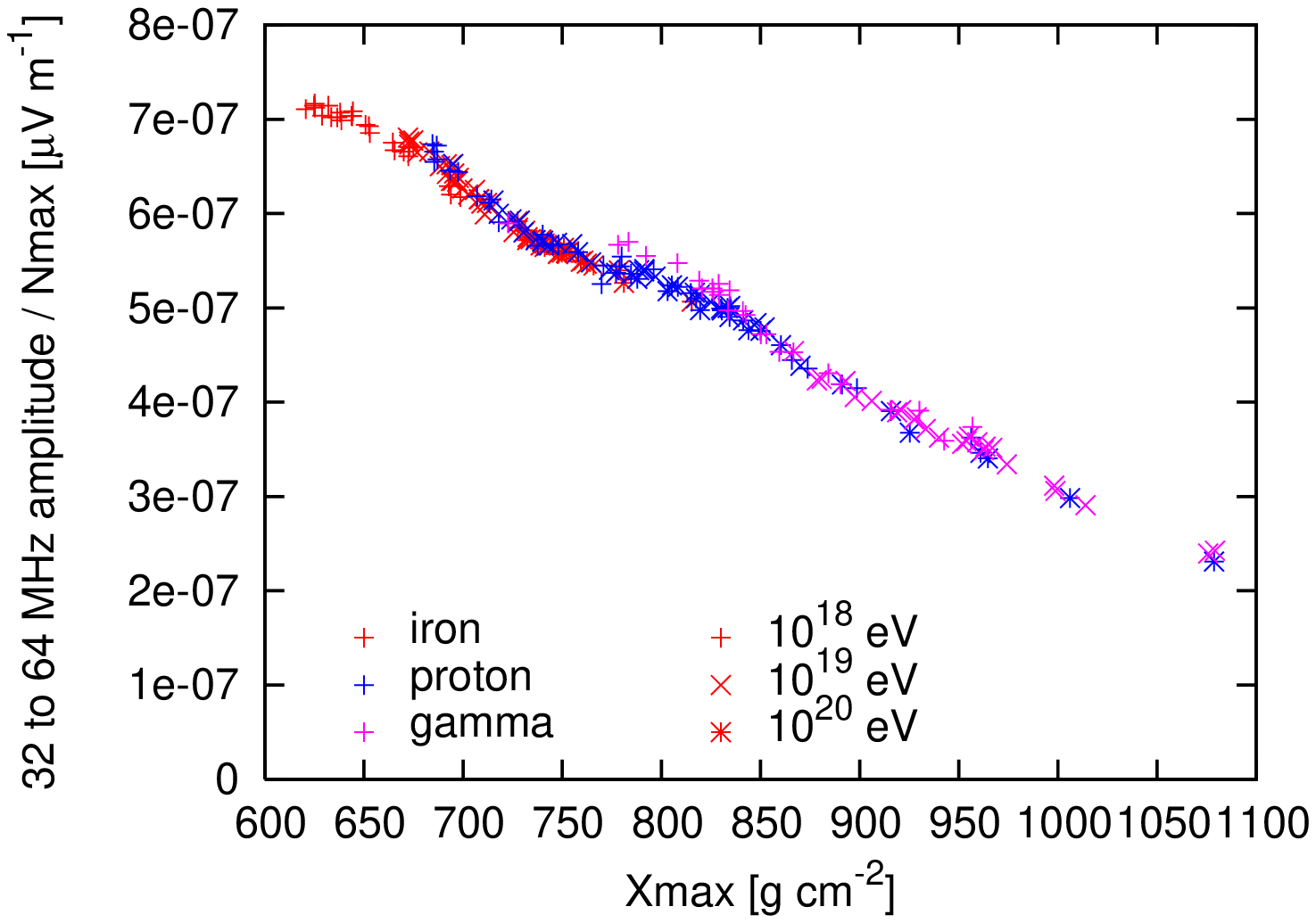}
\caption{In the {\em steep region} at 725~m north, the radio pulse height per $N_{\mathrm{max}}$ over all energies and particle types is constant for a given $X_{\mathrm{max}}$.}\label{fig:steepnorm}
\end{minipage}
\end{figure*}

When the measured radio field strength is divided by the number of electrons plus positrons in $X_{\mathrm{max}}$, hereafter called $N_{\mathrm{max}}$, it becomes clear that the radio signal scales linearly with $N_{\mathrm{max}}$. The reason for this is that most of the radio emission stems from the particles close to the shower maximum \cite{HuegeUlrichEngel2007a}. (The intensity of optical fluorescence light, in contrast, scales with the calorimetric energy deposited in the atmosphere by the shower.) The clean linear scaling is illustrated in Fig.\ \ref{fig:flatnorm}, where the results from all energies and particle species yield an approximately constant electric field strength per $N_{\mathrm{max}}$ in the {\em flat region}. The electric field strength per $N_{\mathrm{max}}$ in the {\em steep region} is also constant over the different particle types and energies for a given $X_{\mathrm{max}}$.

\begin{figure*}[!htb]
\begin{minipage}{0.47\textwidth}
\centering
\includegraphics [width=4.80cm,angle=270]{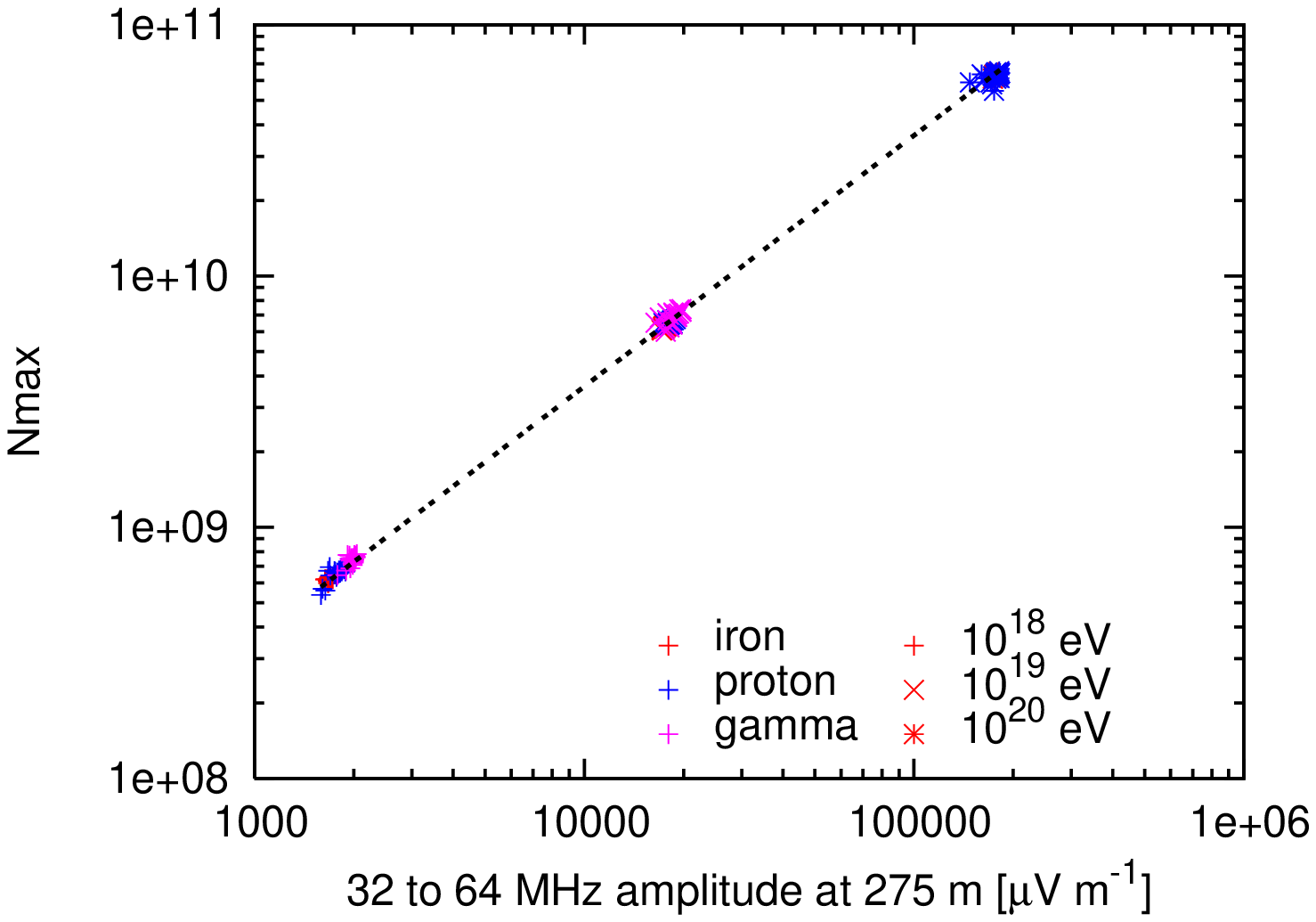}
\caption{A measurement in the {\em flat region} at 275~m north directly yields the shower $N_{\mathrm{max}}$. The line denotes a linear functional dependence.}\label{fig:energy}
\end{minipage}
\hspace{0.05\textwidth}
\begin{minipage}{0.47\textwidth}
\centering
\includegraphics [width=4.80cm,angle=270]{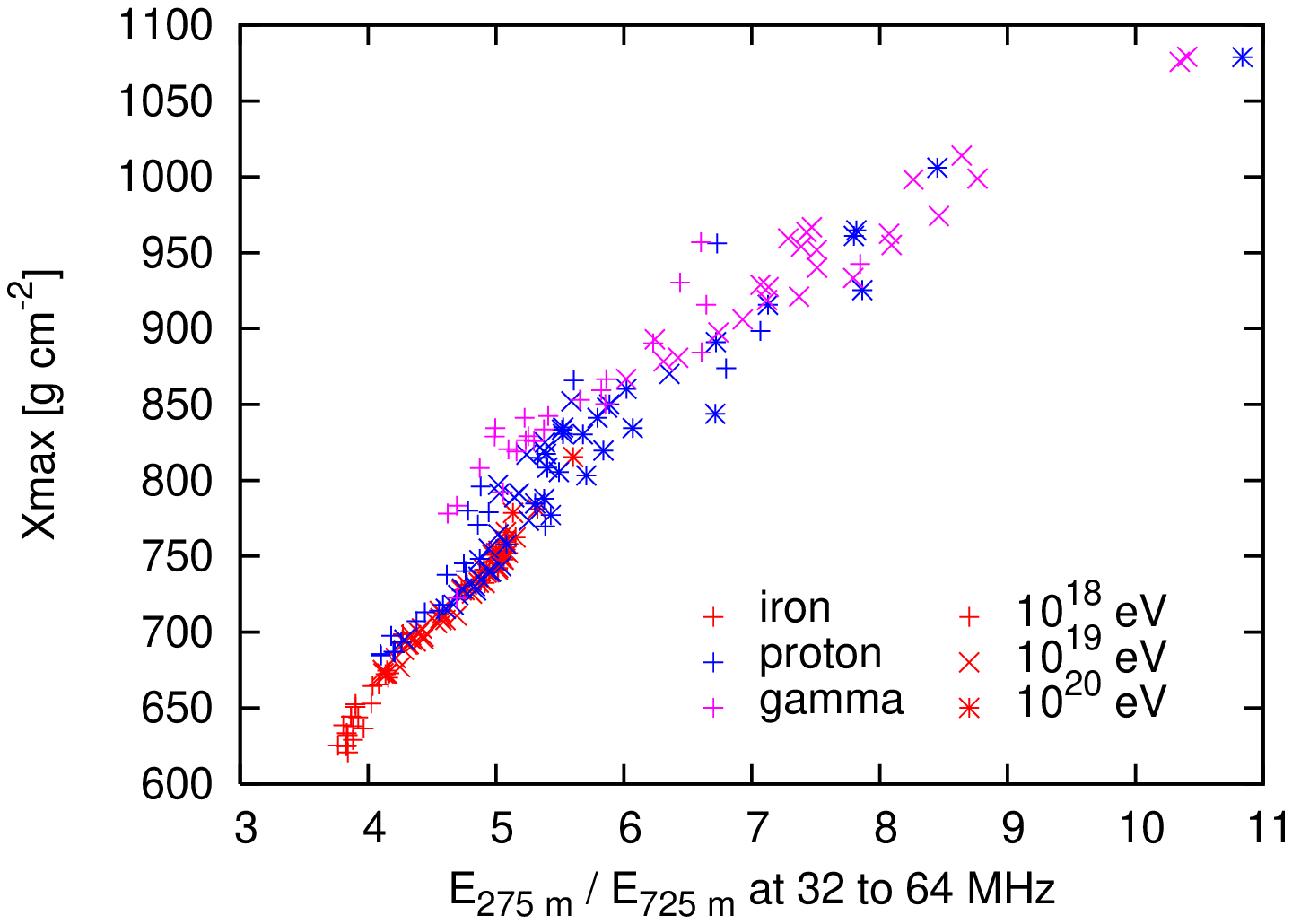}
\caption{The ratio of peak field strengths in the {\em flat region} and {\em steep region} yields direct information on the shower $X_{\mathrm{max}}$.}\label{fig:composition}
\end{minipage}
\end{figure*}

\section{Optimum parameter determination}

\begin{figure*}[!htb]
\begin{minipage}{0.47\textwidth}
\centering
\includegraphics [width=4.80cm,angle=270]{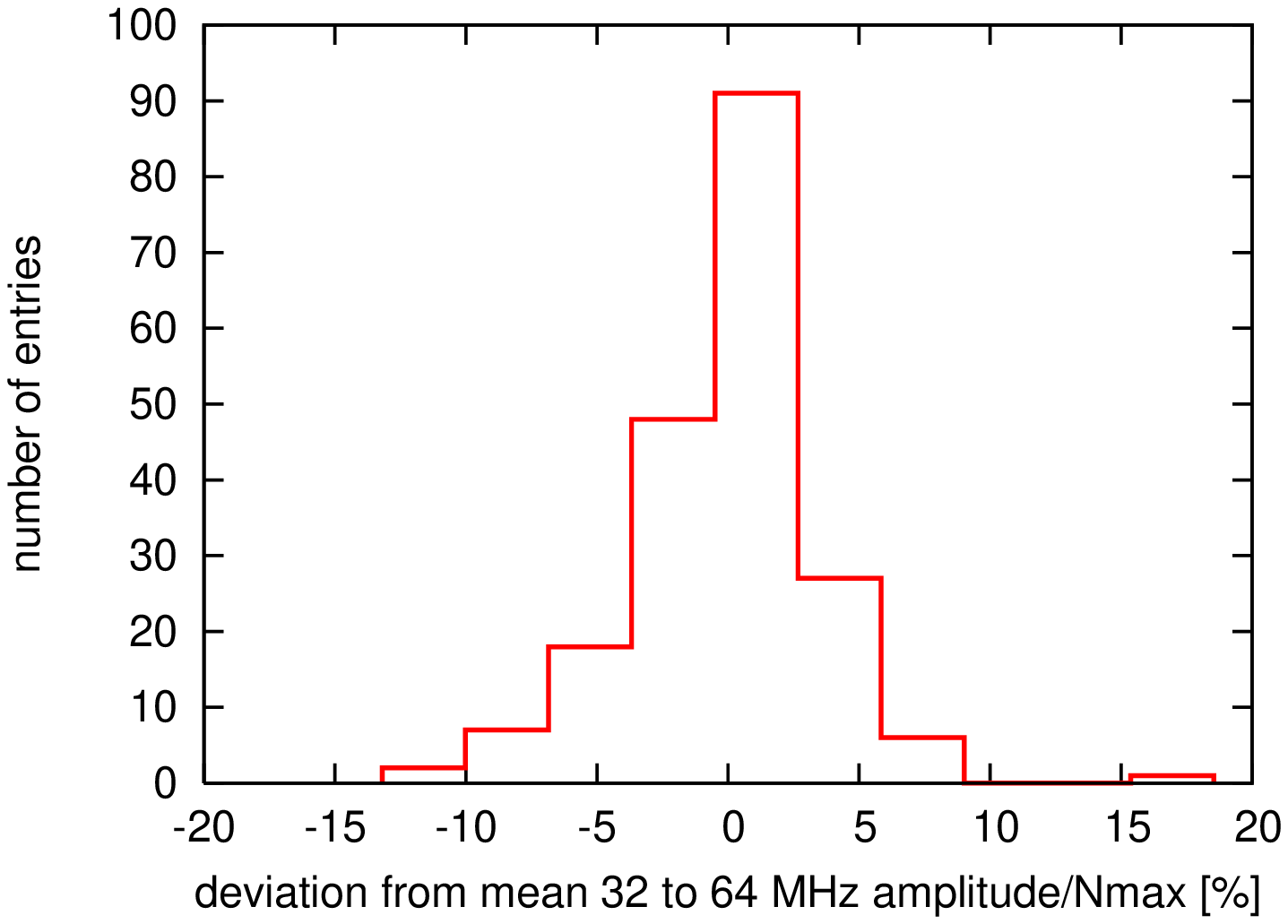}
\caption{For $60^{\circ}$ zenith angle showers, the RMS spread in peak field strength per $N_{\mathrm{max}}$ in the {\em flat region} over all energies and particle types amounts to $\sim5$\%. In principle, $N_{\mathrm{max}}$ can thus be inferred with 5\% uncertainty.}\label{fig:rms}
\end{minipage}
\hspace{0.05\textwidth}
\begin{minipage}{0.47\textwidth}
\centering
\includegraphics [width=4.80cm,angle=270]{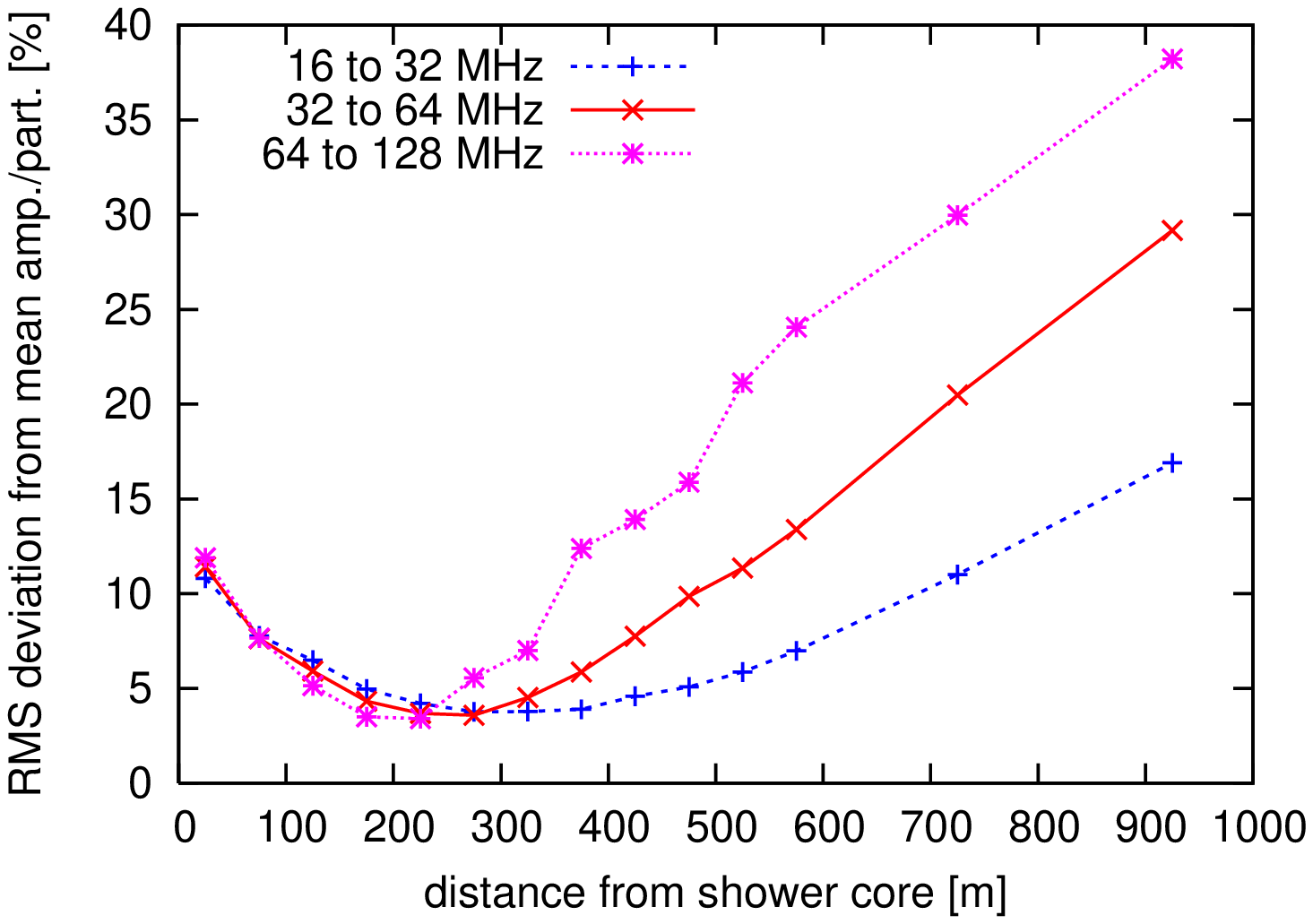}
\caption{Determination of the RMS uncertainty in peak field strength per $N_{\mathrm{max}}$ as a function of lateral (ground-coordinate) distance from the shower core. The {\em flat region} shifts to larger lateral distances at lower frequencies.}\label{fig:rmsdep}
\end{minipage}
\end{figure*}

Histogramming the distribution of electric field strengths per $N_{\mathrm{max}}$ over all particle species and energies yields an RMS of only 5\% in the {\em flat region} (Fig.\ \ref{fig:rms}). In spite of shower to shower fluctuations, radio measurements could thus determine the $N_{\mathrm{max}}$ of individual EAS to very high precision. The position of the {\em flat region} for a given zenith angle is determined by the observing frequency band. As illustrated in Fig.\ \ref{fig:rmsdep}, for $60^{\circ}$ zenith angle and 32 to 64 MHz observing frequency, $N_{\mathrm{max}}$ can be determined to highest precision around 275~m distance to the north (or south). At 16 to 32 MHz, 5\% precision in the $N_{\mathrm{max}}$ determination can be reached anywhere between around 200~m and 500~m from the core. Low frequencies are thus particularly well suited for an energy determination of EAS primaries, if the technical difficulties involved with measurements at these frequencies can be overcome. In contrast, at 64 to 128 MHz the range for $N_{\mathrm{max}}$ determination becomes much smaller and resides at smaller distances. (The peculiar steps in the 64 to 128~MHz RMS curve point to coherence effects starting to play a significant role at these high frequencies.) For the determination of $X_{\mathrm{max}}$ using measurements in the {\em steep region}, a lateral distance should be chosen that shows a large RMS spread in electric field strength per $N_{\mathrm{max}}$. For 32 to 64~MHz and $60^{\circ}$ zenith angle, 725~m constitutes a suitable compromise between a good handle on the $X_{\mathrm{max}}$ value and detectable absolute signal levels. (The absolute signal strength drops quickly with lateral distance, cf.\ \cite{HuegeFalcke2005b}).

\section{Conclusions}

Simulations of geosynchrotron radio emission with the REAS2 Monte Carlo code reveal that information contained in the lateral profile of the radio signal can be exploited for a direct determination of $N_{\mathrm{max}}$ and $X_{\mathrm{max}}$ on a shower-to-shower basis. These quantities can in turn be related to the energy and mass of the primary cosmic ray particle.

For a given zenith angle and observing frequency band, distinct distance regimes denoted here as the {\em flat region} and the {\em steep region} can be identified. Measurements in the {\em flat region} directly yield the $N_{\mathrm{max}}$ of the shower, with an RMS deviation of only 5\%. A comparison of peak field strengths in the {\em steep region} and {\em flat region} for a given observing bandwidth provides a direct estimate of the shower $X_{\mathrm{max}}$. Alternatively, measurements at a fixed distance but in two different observing frequency bands provide comparable information.

The position of the {\em flat region} in case of $60^{\circ}$ zenith angles for the observing frequency band from 32 to 64~MHz lies at approximately 275~m and would thus require an antenna spacing of that order. (The given number is valid for antenna positions along the shower axis; in the perpendicular direction the scales are smaller due to azimuthal asymmetries of the radio footprint.) If technical problems arising in measurements at lower frequencies such as 16 to 32~MHz can be overcome, antenna spacings of up to 500~m will allow measurements in the {\em flat region}. The qualitative behaviour at $45^{\circ}$ zenith angle is analogue, yet at lower lateral distances. At larger zenith angles, the scales will be larger, but consistent simulations require the implementation of curved atmospheres for zenith angles much larger than $60^{\circ}$, which is currently being prepared.

%This is the reference to .bib file (Whitout .bib!)
%\bibliography{tims_references}

\begin{thebibliography}{1}

\bibitem{HeckKnappCapdevielle1998}
D.~{Heck}, J.~{Knapp}, J.~N. {Capdevielle}, G.~{Schatz}, and T.~{Thouw}.
\newblock FZKA Report 6019, Forschungszentrum Karlsruhe, 1998.

\bibitem{HuegeIcrc2005a}
T.~{Huege}, W.~D. {Apel}, A.~F. {Badea} {et al.}
\newblock In {\em Proc. of the 29th ICRC, Pune, India}, volume~7, page 107,
  2005.
\newblock astro-ph/0507026.

\bibitem{HuegeFalcke2003a}
T.~{Huege} and H.~{Falcke}.
\newblock {\em Astronomy \& Astrophysics}, 412:19--34, December 2003.

\bibitem{HuegeFalcke2005b}
T.~{Huege} and H.~{Falcke}.
\newblock {\em Astropart. Phys.}, 24:116, 2005.

\bibitem{HuegeUlrichEngel2007a}
T.~{Huege}, R.~{Ulrich}, and R.~{Engel}.
\newblock {\em Astropart. Physics}, 27:392--405, 2007.

\bibitem{HuegeIcrc2007b}
T.~{Huege}, R.~{Ulrich}, and R.~{Engel}.
\newblock In {\em Proceedings of the 30th ICRC, Merida, Mexico}, 2007.
\newblock (these proceedings).

\bibitem{PetrovicApelAsch2006}
J.~{Petrovic}, W.~D. {Apel}, T.~{Asch}, F.~{Badea} {et al.}
\newblock {\em Astronomy \& Astrophysics}, 462:389--395, 2007.

\end{thebibliography}
%This in the bibtex style, is ok.
%\bibliographystyle{plain}

\end{document}